# Large self-heating by trapped-flux reduction in Sn-Pb solders

Yoshikazu Mizuguchi[1]*, Takumi Murakami[1], Md. Riad Kasem[1], Hiroto Arima[1,2]

1.  *Department of Physics, Tokyo Metropolitan University, Hachioji 192-0397, Japan*

2.  *AIST, 1-1-1 Umezono, Tsukuba, Ibaraki 305-8560, Japan*

(*Corresponding author: mizugu@tmu.ac.jp)

Abstract

Magnetic flux trapping in field-cooled (FC) Sn-Pb solders has been recently studied because of the observation of nonvolatile magneto-thermal switching [H. Arima *et al.*, Commun. Mater. 5, 34 (2024)] and anomalous magnetic field-temperature ($H$-$T$) phase diagrams [T. Murakami et al., AIP Adv. 13, 125008 (2023)]. In this paper, we investigate the origin of the anomalously low specific heat ($C$) in Sn10-Pb90 and Sn45-Pb55 solders after FC at $H$ = 1500 Oe. We show that the FC solders exhibit self-heating possibly caused by the flux flow during the reduction of trapped fluxes when heating the sample during the $C$ measurements. The $T$ dependence of $T$ rise clearly exhibits unexpectedly large values when the low-$C$ states are observed. In addition, the cause of the transition-like behavior in $C$-$T$ of FC solders are explained by local heating during $H$ control and flux-jump phenomena.

Sn-Pb solders are known as phase-separated composite materials with Sn and Pb regions (Fig. 1(a)), both are originally superconducting with transition temperatures ($T_c$s) of 3.7 and 7.2 K, respectively. The Sn-Pb solders trap magnetic field ($H$) when the sample was field-cooled (FC) or experienced $H$ greater than upper critical field of Pb ($H_c$ ~800 Oe at $T$ = 0 K) [1-3]. Recently, we proposed that the trapped fluxes are located at the Sn regions of the solders (Fig. 1(b,c)), and the selective flux trapping achieved nonvolatile magneto-thermal switching, which is a key technology for low-$T$ thermal management [3]. Furthermore, the Sn45-Pb55 (represented as wt% ratio) solder after FC at $H$ = 1500 Oe exhibits anomalous $H$-$T$ phase diagrams, in which $H$-robust lower-$T$ transition is observed in addition to conventional $H$-$T$ for higher-$T$ transitions (superconducting transitions of Pb) [4]. Therefore, understanding of the flux-trapping mechanisms in Sn-Pb solders is an important issue. In Ref. 4, we proposed that the lower-$T$ transitions would be related to superconducting transition of the Sn regions, but the origins of the appearance of the transition-like phenomenon in $C$-$T$ and low-$C$ states have not been clarified. In this paper, we propose that one of the possible causes of the transition-like behavior in $H$-$T$ phase diagram (lower-$T$ transition) in FC solders would be self-heating of the solders caused by the flux-flow heating phenomenon during the $C$ measurements. By checking $T$ dependence of $T$ rise and the waveforms of the $C$ measurements with a relaxation method, we found unexpectedly enhanced $T$ rise for the low-$C$ datapoints. In addition, initial local heating of the sample during $H$ control and flux-jump events result in the suppression of the anomalously high $T$ rise at lower temperatures. The present results do not propose the emergence of superconducting states of the Sn regions at low temperatures [4], but the observation of anomalous self-heating related to the continuous flux reduction depending on $T$ in the solders will be exotic states of composite superconducting materials. Since the amount of the trapped flux of the Sn$x$-Pb(100-$x$) solders increases with decreasing $x$ (Fig. S1), we mainly focus on the properties of $x$ = 10 sample.

The solder samples of Sn$x$-Pb(100-$x$) with $x$ = 10 and 45 are products of Sasaki Solder

Industry and TAIYO ELECTRIC IND. CO., LTD., respectively. The phase separations were investigated by scanning electron microscope (SEM) TM3030 (Hitachi High-Tech) with a backscattering mode. Energy-dispersive X-ray spectroscopy (EDX) for elemental mapping was performed using SwiftED (Oxford) on TM3030. As shown in Fig. 1(a), the μm-scale Sn regions are present in the $x$ = 10 solder. The $T$ and $H$ dependences of $M$ were measured using a superconducting quantum interference device (SQUID) magnetometer on a Magnetic Property Measurement System MPMS3 (Quantum Design). All the $M$ data were corrected by the demagnetization factor. $C$ was measured by relaxation method using a Physical Property Measurement System PPMS-Dynacool (Quantum Design). Apiezon N grease was used to attach the sample to the sample stage of the sample puck.

Fig. 2(a) shows $4\pi M$-$T$ measured at 10 Oe after ZFC, followed by FC at 10 Oe. Conventional diamagnetic signals are observed, and there are no trapped fluxes. Figure 2(b) shows $4\pi M$-$H$ measured at 1.8 K after ZFC. Flux jumps, which are typically observed in superconductors with strong pinning, are observed. $MgB_2$ is a typical system where the flux jumps are observed [5-7], and nonvolatile magneto-thermal switching was also observed in $MgB_2$ [8]. Therefore, the pinning of Sn-Pb solders is also strong and affects physical properties.

Fig. 3(a) shows $4\pi M$-$T$ measured without external field after FC (1500 Oe). Because of the magnetic flux trapping, no external field is needed for taking the $M$ data. Similar measurements have been performed in hydride superconductors [9], and Minkov et al. proposed the presence of strong flux pinning in the hydrides. In $H_3S$ [9], the $T$ cycles after FC produced flat (independent of temperature) and tunable magnetization. Similar $4\pi M$-$T$ curves were obtained for the Sn10-Pb90 solder by the various $T$ cycles after FC (at 1500 Oe) as shown in Fig. 3(b). These facts also suggest the magnetic fluxes trapped in the solders are strongly pinned. Figures 3(c) and 3(d) show the $4\pi M$-$T$ measured at various fields after FC (1500 Oe). For the cases with positive fields (same direction as the FC field), $4\pi M$ constantly decreases with increasing external field, which is caused by the compensation of inner magnetic field in the solders where the change in inner magnetic field is preferred to be maintained after the external field exposure. For the cases with negative fields (opposite direction as the FC field), the lowest-$T$ value of $4\pi M$ increases by the increase in the absolute value of the negative fields until $H$ = -500 Oe. This trend can also be understood by the compensation scenario, and the $4\pi M$ is enhanced by the negative $H$, which means that the trapped fields are preserved. As shown in Fig. S2, the $B$-$T$ curves under most of the applied fields exhibit the similar curve, which supports the above scenario. As seen for several data with negative fields, jumps are observed in $4\pi M$-$T$. The jump should be related to the flux jumps observed in $4\pi M$-$H$ (Fig. 2(b)). At $H$ < -500 Oe, the amount of trapped field decreased, which is due to the weakened supercurrent of Pb by the fields.

Next, we briefly discuss the $C$ data taken after FC (1500 Oe). Here, one measurement was performed at each $T$ after heating the sample to the target $T$ (see following discussion part for multiple-measurement results). First, normal-state $C$-$T$ was measured at $H$ = 1000 Oe, and the electronic-specific-heat coefficient ($\gamma$) was estimated from the fitting of $C(T, H = 1000$ Oe$)$ to the equation of low-$T$ limit, $C(T) = \gamma T + \beta T^3$ as shown in Fig. S3. The estimated $\gamma$ is 2.27(8) mJK$^{-2}$mol$^{-1}$. As performed in Ref. 4, to investigate the contributions related to superconducting states, the $\delta C/T$ data were estimated by subtracting the normal-state $C(T, H = 1000$ Oe$)/T$ data (measured at $H$ = 1000 Oe) from the total specific heat: $\delta C/T = [C(T, H) - C(T, H = 1000$ Oe$)]/T$. To visually compare the evolutions of the transitions, the obtained $\delta C/T$ data are normalized and plotted in arbitrary unit in Fig. S4(a). From the anomaly $T$, an $H$-$T$ phase diagram was created as shown in Fig. S4(b). Although the lower-$T$ transitions look superconducting transitions, the estimated $\delta C/T$ data exhibits quite large jump, which cannot be explained by electronic entropy change of Sn within the conventional model [10]. In Fig. S5, the $T$ dependences of $\delta C/T$ measured at $H$ = 0 and -500 Oe after FC (1500 Oe) are shown. For example, $\Delta(\delta C/T)$ for the lower-$T$ transition, $\Delta(\delta C/T)_{\text{low}}$, is 4.66 mJK$^{-2}$mol$^{-1}$. With $\gamma$ (total),

Δ(δ$C/T$)$_{low}$/$γ$ (total) is 2.05, which is greater than the BCS value expected for pure Sn, and, in addition, this would be overestimated because of the use of $γ$ (total). Furthermore, the Δ(δ$C/T$)$_{low}$/$γ$ (total) is clearly enhanced with the applied fields of $H$ < 0, and from the data measured at $H$ = -500 Oe, Δ(δ$C/T$)$_{low}$/$γ$ (total) = 3.96 is obtained. When calculating Δ(δ$C/T$)/$γ$ (Sn) with the weighted $γ$ (Sn) estimated from molar ratio of Sn and Pb, Δ(δ$C/T$)$_{low}$/$γ$ (Sn) is estimated as 12.63 and 24.35 for the data at $H$ = 0 Oe and -500 Oe, respectively. Therefore, the lower-$T$ transitions cannot be understood by the electronic entropy change. Furthermore, we noticed that the $C$ values between the lower-$T$ transition and $T_c$ of Pb are anomalously low. Hence, we took multiple datapoints at each T and found that only the first datapoint shows the extremely-low $C$ and the second and third datapoints become higher.

In Figs. 4(a) and 4(b), $C$-$T$ for $x$ = 10 measured at $H$ = 0 Oe (after ZFC) and $H$ = 1500 Oe (> $H_c$) are displayed. In Figs. 4(c) and 4(d), $C$-$T$ for $x$ = 10 measured at $H$ = 0 and -500 Oe (both after FC (1500 Oe)) are displayed. As shown in Fig. 4(c), the $C$ datapoints fluctuate above 2 K, and the first data after reaching (heating the sample to) the target $T$ is low, and the second and third data are almost same and high. Below 2 K, this behavior is not observed. Similar behavior is seen in Fig. 4(d) at $T$ > 3 K. For ZFC data and normal-state ($H$ = 1500 Oe) data (Figs. 4(c) and 4(d)), the data fluctuation is not observed, which means that the behavior is unique for FC sample with flux trapping. To understand the cause of the appearance of low-$C$ states, the $T$-time waveforms in the relaxation method were carefully checked. Then, we found that the $T$ rise, which was set to 2%, becomes clearly larger than 2% in the data-fluctuation $T$ regime. The typical waveforms are shown in Fig. 5. For first datapoint, additional heating is seen after the initial heating in the $C$ measurement. The red curve in Fig. 5 is the extreme case with the resulting $T$ rise exceeding 3%. The green curve shows the normal case with normal-conducting states at $H$ > $H_c$; no anomaly is seen during heater heating. However, for example, the experimental waveform with a $T$ rise of ~2.5%, which results in low-$C$ data as well, does not show a clear anomaly during heater heating (deviation from the fitting curve). Namely, checking the deviation of $T$ rise from the expected values and the difference between first and second (and third) datapoints are the way to detect self-heating possibly caused by flux motion. In Figs. 4(e)-4(h), the corresponding plots of the $T$ dependence of $T$ rise are displayed. There are no fluctuations of $T$ rise in Figs. 4(e) and 4(f), but the fluctuations are observed in Figs. 4(g) and 4(h) for the FC sample with trapped fluxes. This difference and the fluctuation of $T$ rise would be caused by self-heating of the sample, which is related to the decrease in trapped field by heating during the $C$ measurements. When the $T$ rise was greater than that expected from the given energy from the sample heater, the resulting $C$ is directly affected and becomes larger. This assumption explains the reason why only the first datapoint is affected because no reduction of trapped field occurs (see magnetization part later) when the sample $T$ is lower than the highest $T$ experienced in the first measurement at the $T$. As a fact, in Fig. 5, the activation of the additional heating is seen at $T > 2.225$ K and $t > 0.7$ s; this is caused by the experience of $T \sim 2.225$ K in the previous experiments just before heating to this target $T$. Therefore, we consider that the self-heating by flux flow due to the reduction of the trapped field and its instability [11-15] only occur in the first measurement after reaching the target $T$.

Next, we investigated the effects of $H$ control rate; after FC, $H$ was controlled with a rate of 200 Oe/s or 10 Oe/s to the target $H$, which should affect sample and/or sample-puck heating due to Eddy current. For the data shown in Figs. S4 and 4, $H$ was controlled with a rate of 200 Oe/s, and in Fig. 6, similar multiple measurements were performed after $H$ control with a rate of 10 Oe/s to avoid heating of sample and/or sample puck. As shown in Fig. 6(a), the anomalous self-heating data are observed at whole $T$ range for $H = 0$ Oe after FC. The results suggest that the high $H$ control rate of 200 Oe/s results in a $T$ increase of ~0.2 K. Similar data was observed for $H = -400$ Oe (Fig. 6(b)), but for $H = -500$ Oe (Fig. 6(c)), the anomalous self-heating is not observed below 3 K. The difference would be understood by considering the flux jump. As shown in Fig. 3(d), flux jump is not observed for $H \geq -400$ Oe. Therefore, if the initial sample heating can be ignored, the flux reduction occurs from the lowest $T$ for $H ≥ -400$ Oe. However, with $H ≤ -500$ Oe and/or initial sample heating with a high $H$ control rate, a regime with no flux reduction can be created after FC process, which results in the results like Fig. 6(c). Similar anomalous heating was observed in a Sn45-Pb55 solder sample as shown in Fig. S6.

We assume that the observed self-heating in the Sn-Pb solders after FC is caused by flux motions. Therefore, the trapped fluxes would be present in a form of vortex, and the motion of vortex would result in local Joule heating. Flux jumps and/or flux motion have been observed in the mixed states of type-II superconductors [5-7, 11-19], but vortex can also be hosted by the intermediate state of type-I superconductors [20-22]. Furthermore, in Ref. 19, the simulation results showed motion of vortices in the intermediate states. Hence, we consider that the fluxes are trapped in the Sn regions in the vortex form, and the motion of vortices causes sample heating for the first measurement at each target $T$. To clarify the details on the form of fluxes in solders, direct observation of vortices using local probes are desired.

Next, we briefly discuss about the possible causes of self-heating phenomenon in the FC solders. The flux jumps or vortex motions in the above-mentioned systems are generally induced by the change in external fields like magnetic field or electric current. In addition, by heating the sample to close to $T_c$, flux-flow resistance is observed, and the relation to self-heating has been theoretically demonstrated [23–25]. In the present solder case, the heating is observed in a relatively wide temperature range without significant perturbation. For example, as shown in Figs. 6(d) and 6(e), the self-heating is observed from the lowest $T$ (~1.9 K) to 4 K for $H = 0$ Oe and 5 K for $H = -400$ Oe. Since $T_c$ of Sn and Pb are 3.7 K and 7.2 K, respectively, the $T$ range where the flux-flow heating is observed is clearly wide and not limited to the vicinity of critical points like $H_c$ or $T_c$. In the cause of flux jumps [26], heating occurs at $T$ lower than $T_c$ with a clear reduction of $M$ [26], which is seen in Figs. 2(b) and 3(d). However, the self-heating of solders also occurs without reduction of $M$. These facts suggest that the heating cannot be fully explained by flux flow and/or flux jump. Because of the phase separated state and the components of the Sn-Pb solders, which are originally type-I superconductors Sn and Pb, there is possibility of the creation of intermediate state [27]. The configuration of the fluxes in the normal-conducting regions in the intermediate states depend on the magnetic field and morphologies of the sample [27]. Therefore, with decreasing trapped fluxes, large motion of the fluxes may occur. However, to conclude the type of flux motion and the microscopic states of the fluxes, direct observation of trapped fluxes is highly desired. The additional increase in $T$ during the $C$ measurements is sometimes comparable to the target $T$ rise. Therefore, we expect that the slight addition of heat results in the reduction of flux and local heating, and the flux motion should continuously occur at least up to $T$ where the additional $T$ rise is observed. In other words, the FC Sn-Pb solders can spontaneously generate heat by reducing flux trapped in the Sn regions. During the heating, no chemical reaction occurs, and no compositional change emerges. This interesting phenomenon in a flux-trapped composite superconductor may open new thermal-management strategy for device application. For that, experimental investigation on the self-heating in nearly isolated Sn-Pb solder samples is desired.

In conclusion, in Ref. 4, we reported anomalous $H$-$T$ phase diagram for FC Sn-Pb solders and anomalously low-$C$ state in the $C/T$-$T$ data taken in the single-scan mode. In this study, we investigated the causes of the lower-$T$ transition-like behavior and the low-$C$ states by measuring $M$ and $C$ of Sn10-Pb90 and Sn45-Pb55 solders under various $T$ and $H$ control sequences. From $M$ data, strong pinning states are confirmed. From multiple measurements (3 times at each $T$) of $C$, we observed that the low-$C$ state is solely observed in the first datapoints, and high-$C$ data are obtained for the second and third datapoints. In addition, self-heating phenomenon was observed in the first measurement only. These facts suggest that the flux-flow heating occurs when the trapped field is reduced by heating the Sn-Pb solder sample. The additional self-heating results in the low-$C$ data. In addition, the lower-$T$ transition-like behavior was explained by the combination of the absence of the self-heating effects caused by initial sample heating during $H$ control with a high rate and flux jump. Although the current work denies the presence of a lower-$T$ phase transition in the FC Sn-Pb solders, the observation of large self-heating in a wide $T$ range will be unique for this system and will open new pure and applied science of composite superconductors with flux trapping.

**Acknowledgements**

The authors thank O. Miura, M. Ichioka, N. L. Saini, E. Arahata, H. Mori, K. Uchida, F. Ando, H. Sepehri-Amin, and T. Yagi for supports in research and fruitful discussion. This work was partly supported by JST-ERATO (JPMJER2201) and TMU Research Project for Emergent Future Society.

**Data availability**

All the experimental data can be provided by request to the corresponding author.

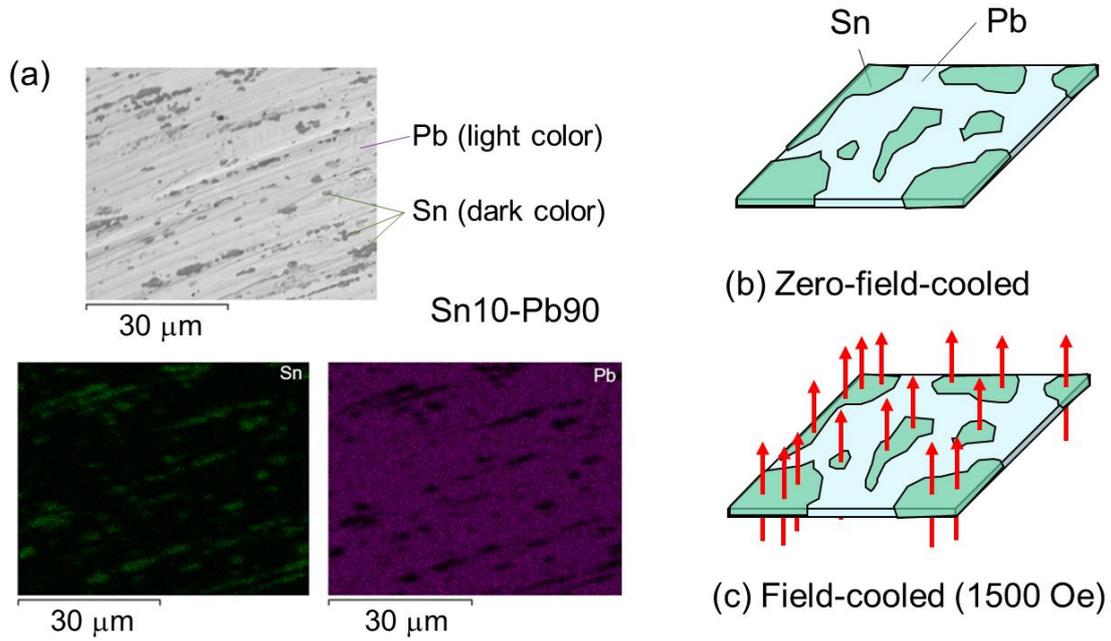

Fig. 1. (a) SEM image and EDX mapping of Sn and Pb for the Sn10-Pb90 ($x = 10$) solder. (b,c) Schematic images of expected flux trapping in ZFC and FC solders.

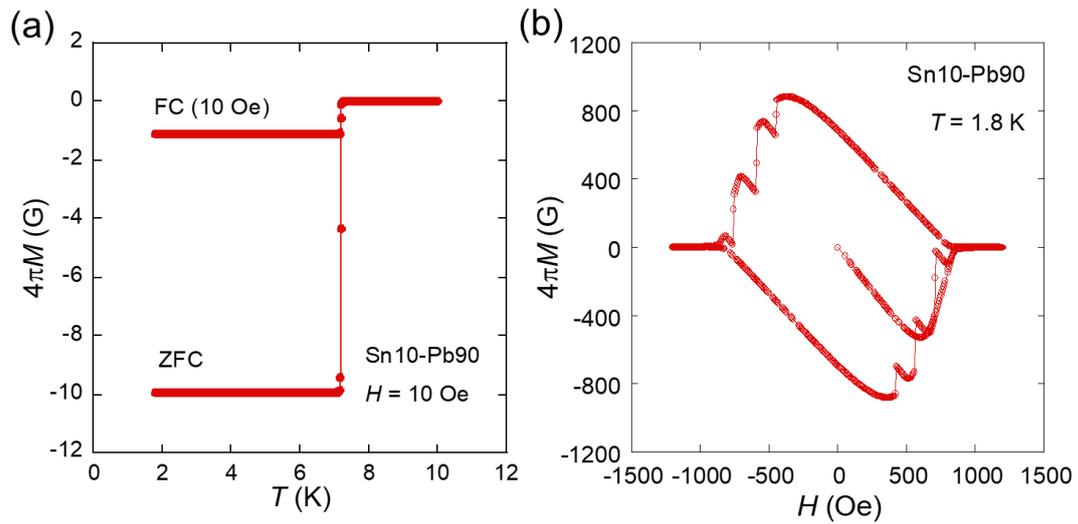

Fig. 2. (a) $T$ dependence of $4\pi M$ for $x = 10$ measured at $H = 10$ Oe after ZFC and FC (10 Oe). (b) $H$ dependence of $4\pi M$ for $x = 10$ measured after ZFC.

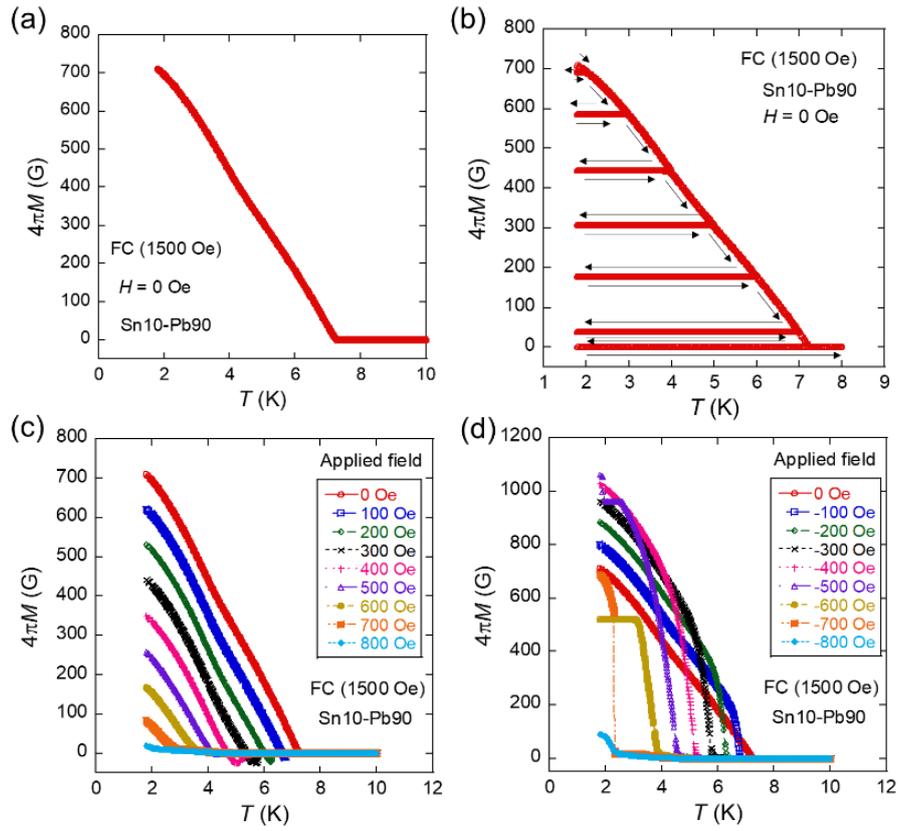

Fig. 3. (a) $T$ dependence of $4\pi M$ for $x = 10$ measured at $H = 0$ Oe after FC (1500 Oe). (b) $T$ evolution of $4\pi M$ measured at $H = 0$ Oe after FC (1500 Oe) with the $T$ sequence as indicated by the arrows. (c,d) $T$ dependence of $4\pi M$ for $x = 10$ measured at various $H$ after FC (1500 Oe). The $H$ control rate was 50 Oe/s.

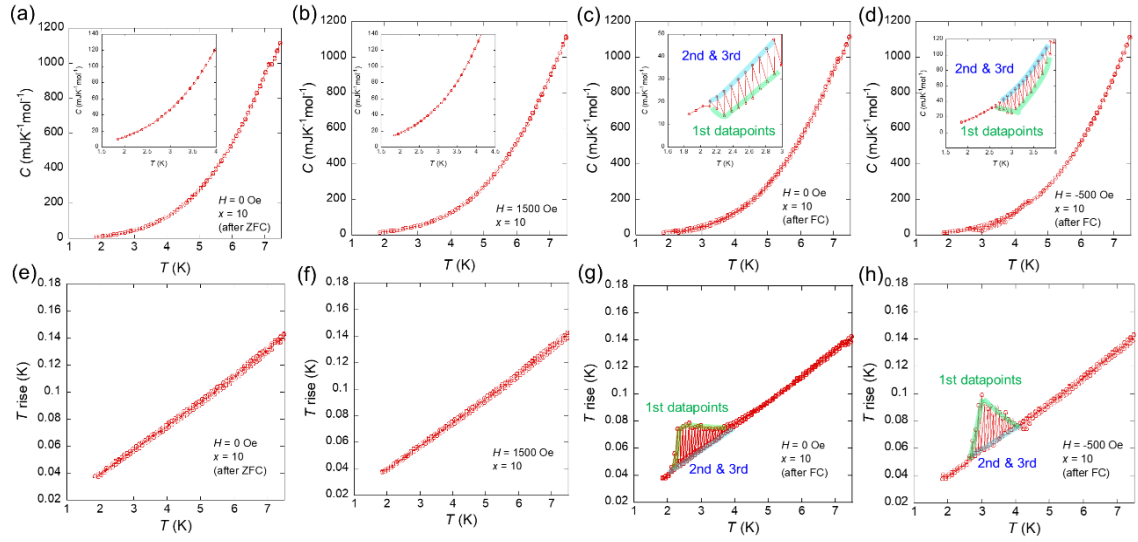

Fig. 4. (a-d) $T$ dependences of $C$ for $x = 10$ measured at (a) $H = 0$ Oe after ZFC, (b) $H = 1500$ Oe, (c) $H = 0$ Oe after FC (1500 Oe), and (d) $H = -500$ Oe after FC (1500 Oe). The $H$ control rate was 200 Oe/s. (e-h) $T$ dependence of $T$ rise of the $C$ measurements in Fig. 4(a-d), respectively. For all the measurements, the target $T$ rise was set as 2%.

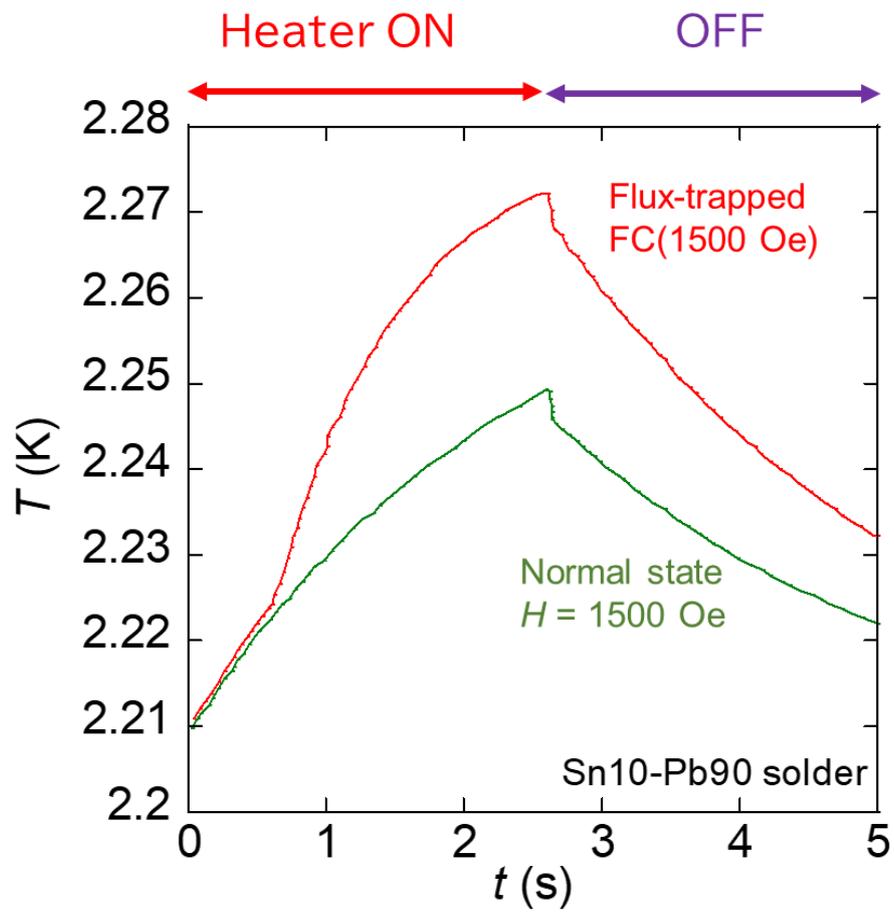

Fig. 5. Waveforms of specific heat measurements (red) with additional self-heating (larger $T$ rise than the target $T$ rise of 2%) possibly caused by the flux motion (reduction) and (green) with normal $T$ rise without flux reduction for Sn10-Pb90.

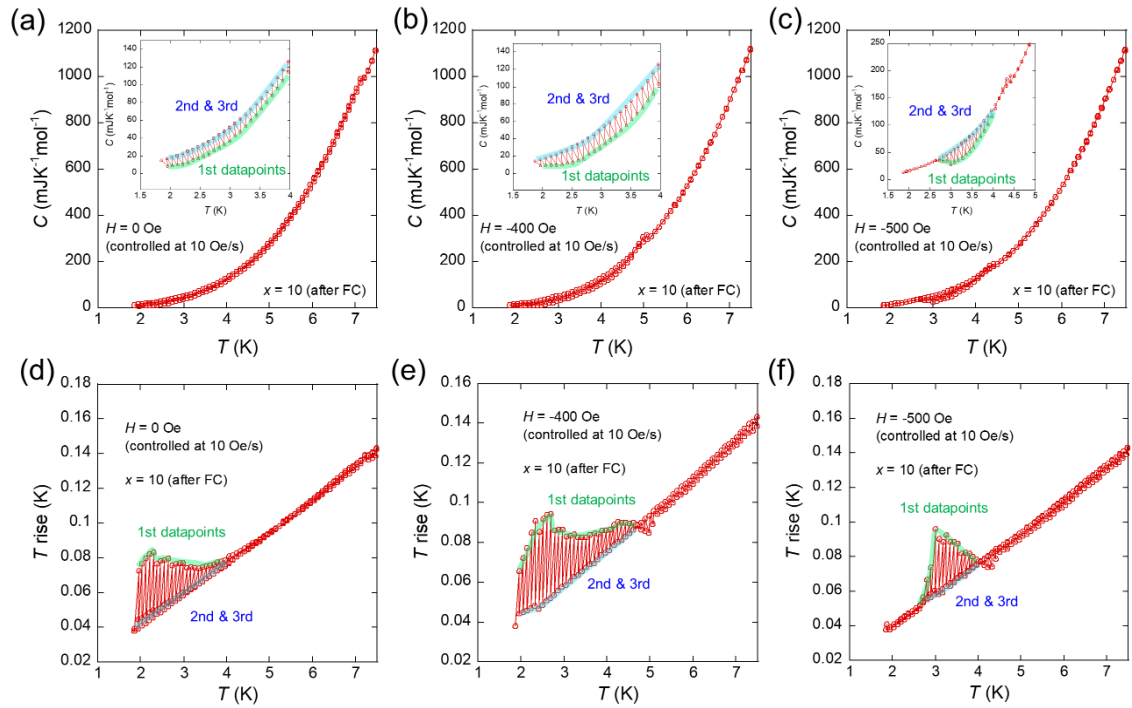

Fig. 6. (a-c) $T$ dependences of $C$ for $x = 10$ measured at $H = 0, -400, -500$ Oe after FC (1500 Oe) with an $H$ control rate of 10 Oe/s. (d-f) $T$ dependence of $T$ rise of the $C$ measurements in Fig. 6(a-c), respectively. For all the measurements, the target $T$ rise was set as 2%.

# Supplementary material

# Large self-heating by trapped-flux reduction in Sn-Pb solders

Yoshikazu Mizuguchi[1]*, Takumi Murakami[1], Md. Riad Kasem[1], Hiroto Arima[1,2]

1. *Department of Physics, Tokyo Metropolitan University, Hachioji 192-0397, Japan*

2. *AIST, 1-1-1 Umezono, Tsukuba, Ibaraki 305-8560, Japan*

(*Corresponding author: mizugu@tmu.ac.jp)

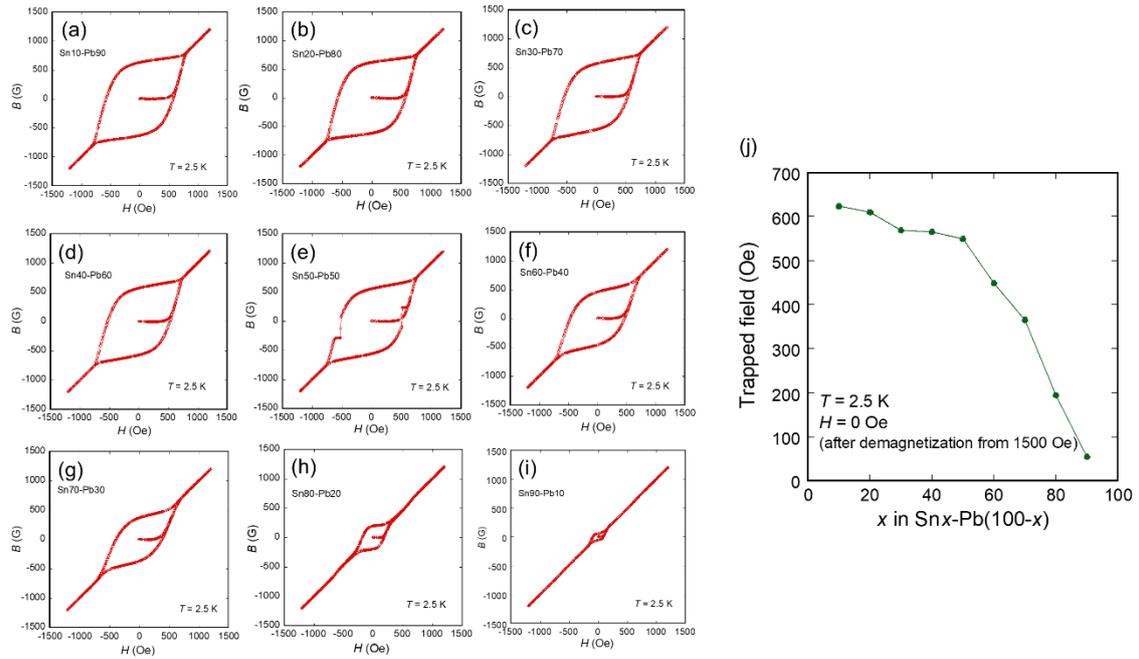

Fig. S1. (a-i) $H$ dependence of inner magnetic field ($B$) at $T$ = 2.5 K for Sn$x$-Pb(100-$x$). (j) Trapped field plotted as a function of Sn mass ratio ($x$). The solder samples of Sn$x$-Pb(100-$x$) with $x$ = 10, 20, 30, 40, 50, 60, 70, 80, 90 were purchased from Sasaki Solder Industry; the typical purities are 5N for $x$ = 10, 20 and 3N for $x$ = 90.

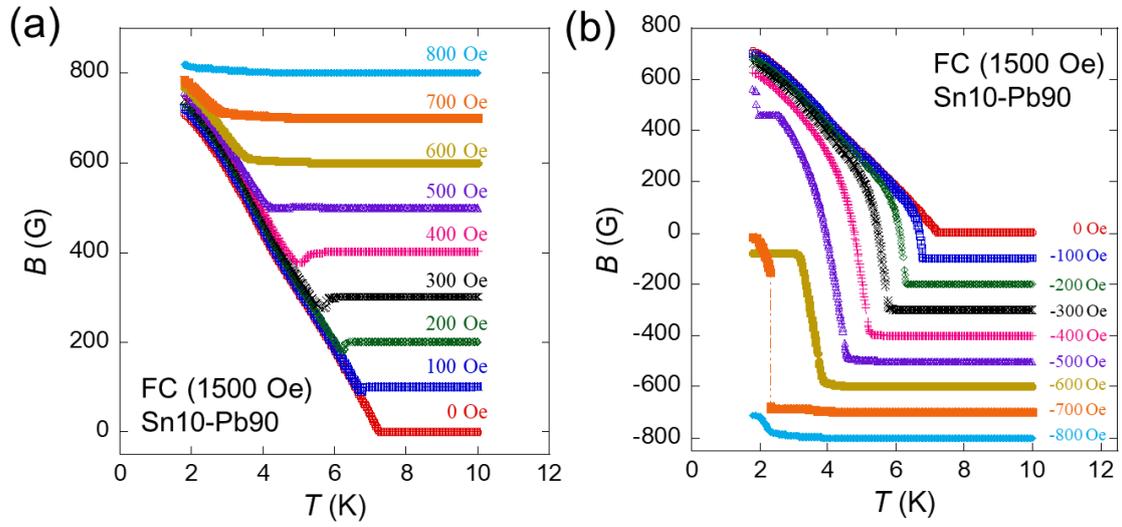

Fig. S2. Temperature dependences of inner magnetic field $B$ (= $4\pi M + H$) for Sn10-Pb90.

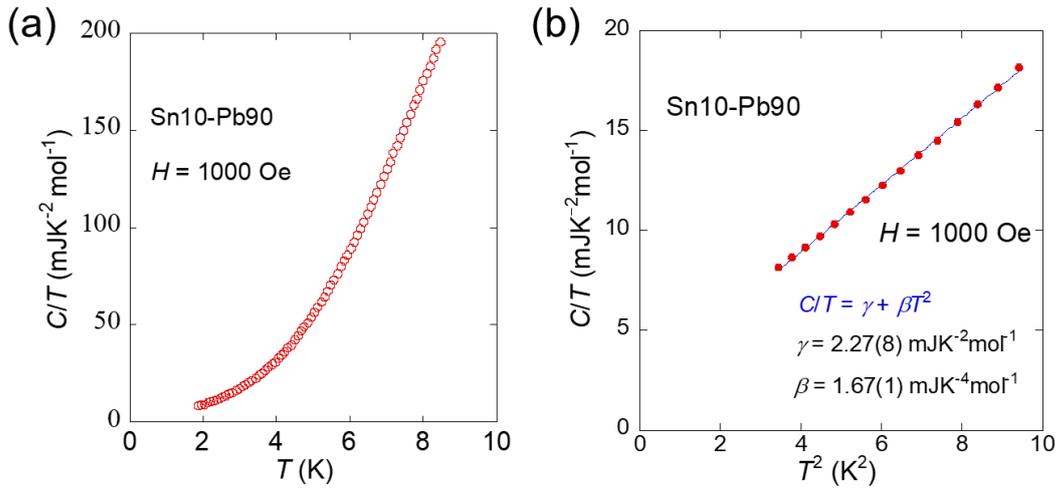

Fig. S3. (a) $T$ dependence of $C/T$ measured at 1000 Oe for Sn10-Pb90. (b) $C/T$-$T^2$ plot of the same data as Fig. S3(a) and fitting result.

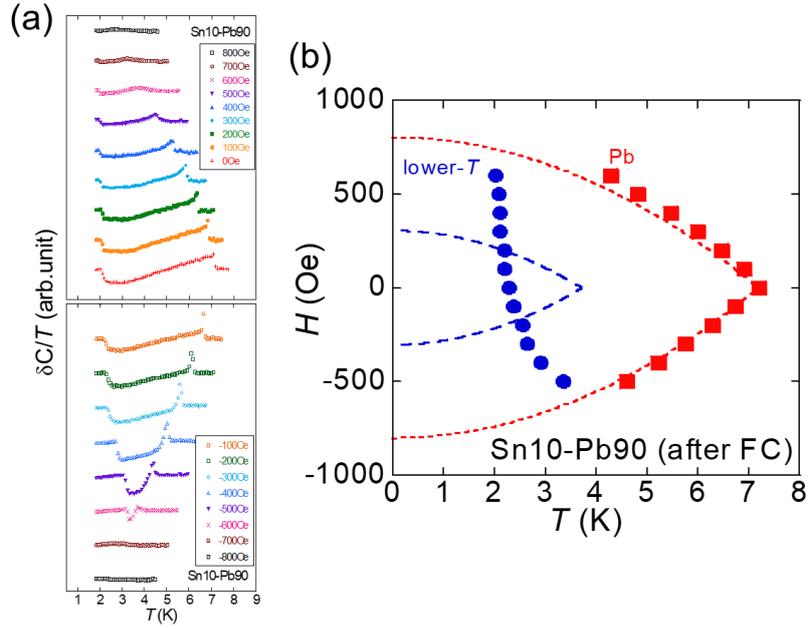

Fig. S4. (a) $T$ dependence of $\delta C/T$ for $x = 10$ measured after FC (1500 Oe). (b) $H$-$T$ phase diagram made by taking the anomaly temperatures. The dushed lines show theoretical values of $H_c$s for Sn and Pb.

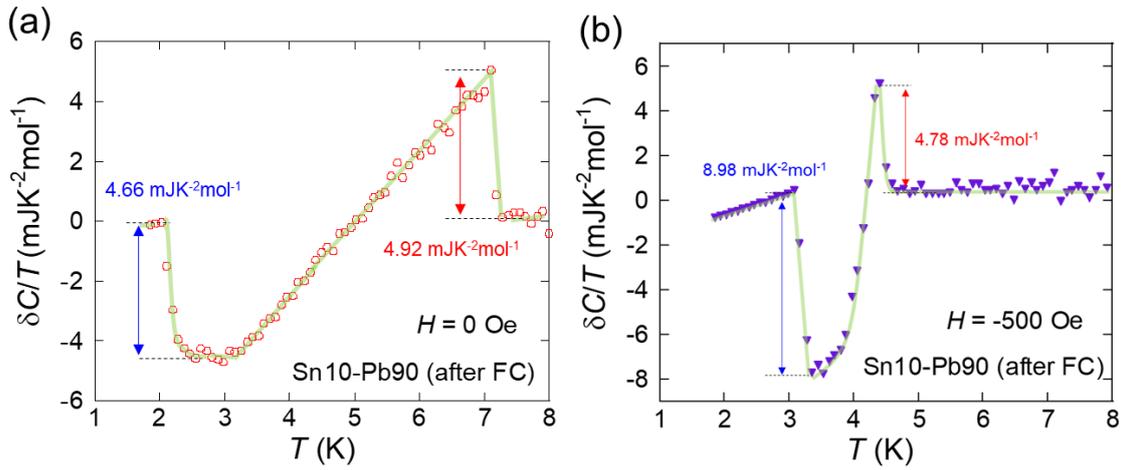

Fig. S5. $T$ dependences of $\delta C/T$ measured at (a) $H = 0$ Oe after FC (1500 Oe) and (b) $H = -500$ Oe after FC (1500 Oe). The light-green lines are eye-guide, and the jump magnitudes are roughly estimated.

Here, we show the results for $x = 45$ measured with the conditions similar to Fig. 6 of the main text. Figure S6(a) shows $C$-$T$ for $x = 45$ measured at $H = 0$ Oe after FC (1500 Oe) with an $H$ control rate of 10 Oe/s. Although the fluctuation magnitude is smaller than that for $x = 10$ (Fig. 6 of the main text), low-$C$ data are obtained for the first datapoint, and the $C$ for the second and third datapoints are higher and comparable. Figure S6(b) shows the $T$ dependence of $T$ rise for $x = 45$. Therefore, the low-$C$ data and the transition-like behavior reported in Ref. S1 can be explained by the above-mentioned scenario with self-heating caused by the flux flow and initial sample heating by a high $H$-control rate.

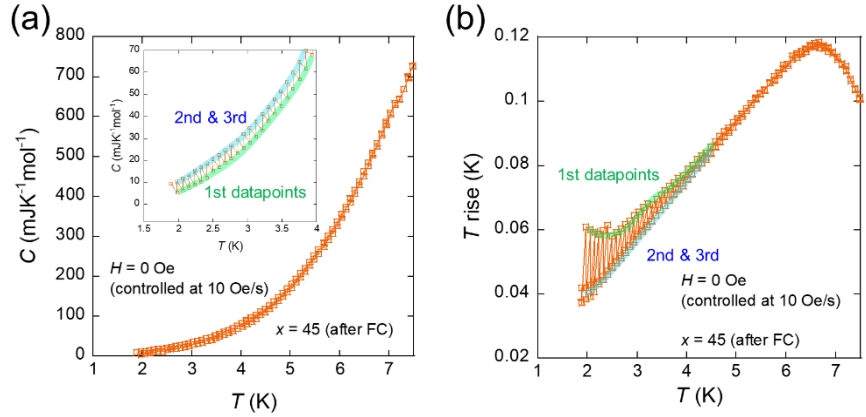

Fig. S6. (a) $T$ dependences of $C$ for $x = 45$ measured at $H = 0$ Oe after FC (1500 Oe) with an $H$ control rate of 10 Oe/s. (b) $T$ dependence of $T$ rise of the $C$ measurements in Fig. 6(a). For all the measurements, the target $T$ rise was set as 2%.

[S1] T. Murakami, H. Arima, and Y. Mizuguchi, AIP Adv. 13, 125008 (2023).20